\documentclass[twocolumn, twocolappendix]{aastex63}

\newcommand{\gb}[1]{\textcolor{blue}{}}
\newcommand{\db}[1]{\textcolor{teal}{}}

\received{November 26, 2021}
\revised{January 13, 2022}
\accepted{January 18, 2022}
\submitjournal{ApJ}

\shorttitle{XTE~J1810$-$197 at submillimeter wavelengths}
\shortauthors{Torne et al.}

\graphicspath{./}

\usepackage{upgreek}

\begin{document}

\title{Submillimeter pulsations from the magnetar XTE~J1810$-$197}

\correspondingauthor{Pablo Torne}
\email{torne@iram.es}

\author[0000-0001-8700-6058]{Pablo Torne}
\affiliation{Institut de Radioastronomie Millim\'etrique, Avda. Divina Pastora 7, Local 20, 18012 Granada, Spain}
\affiliation{East Asian Observatory, 660~N.~A`oh\={o}k\={u}~Place, Hilo, Hawaii, 96720, USA}
\affiliation{Max-Planck-Institut f\"{u}r Radioastronomie, Auf dem H\"{u}gel 69, D-53121, Bonn, Germany}

\correspondingauthor{Graham Bell}
\email{g.bell@eaobservatory.org}
\author[0000-0003-0438-8228]{Graham S. Bell}
\affiliation{East Asian Observatory, 660~N.~A`oh\={o}k\={u}~Place, Hilo, Hawaii, 96720, USA}

\author{Dan Bintley}
\affiliation{East Asian Observatory, 660~N.~A`oh\={o}k\={u}~Place, Hilo, Hawaii, 96720, USA}

\author[0000-0003-3922-4055]{Gregory Desvignes}
\affiliation{LESIA, Observatoire de Paris, Université PSL, CNRS, Sorbonne Université, Université de Paris, 5 place Jules Janssen, 92195 Meudon, France}
\affiliation{Max-Planck-Institut f\"{u}r Radioastronomie, Auf dem H\"{u}gel 69, D-53121, Bonn, Germany}

\author[0000-0001-6524-2447]{David Berry}
\affiliation{East Asian Observatory, 660~N.~A`oh\={o}k\={u}~Place, Hilo, Hawaii, 96720, USA}

\author[0000-0002-5457-9025]{Jessica T. Dempsey}
\affiliation{East Asian Observatory, 660~N.~A`oh\={o}k\={u}~Place, Hilo, Hawaii, 96720, USA}

\author[0000-0002-3412-4306]{Paul T. P. Ho}
\affiliation{East Asian Observatory, 660~N.~A`oh\={o}k\={u}~Place, Hilo, Hawaii, 96720, USA}
\affiliation{Academia Sinica Institute of Astronomy and Astrophysics, P.O. Box 23-141, Taipei 106, Taiwan}

\author[0000-0002-6327-3423]{Harriet Parsons}
\affiliation{East Asian Observatory, 660~N.~A`oh\={o}k\={u}~Place, Hilo, Hawaii, 96720, USA}

\author[0000-0001-6196-4135]{Ralph~P.~Eatough}
\affiliation{National Astronomical Observatories, Chinese Academy of Sciences, 20A Datun Road, Chaoyang District, Beijing 100101, PR China}
\affiliation{Max-Planck-Institut f\"{u}r Radioastronomie, Auf dem H\"{u}gel 69, D-53121, Bonn, Germany}

\author[0000-0002-5307-2919]{Ramesh Karuppusamy}
\affiliation{Max-Planck-Institut f\"{u}r Radioastronomie, Auf dem H\"{u}gel 69, D-53121, Bonn, Germany}

\author[0000-0002-4175-2271]{Michael Kramer}
\affiliation{Max-Planck-Institut f\"{u}r Radioastronomie, Auf dem H\"{u}gel 69, D-53121, Bonn, Germany}
\affiliation{Jodrell Bank Centre for Astrophysics, School of Physics and Astronomy, The University of Manchester, Manchester M13 9PL, UK}

\author[0000-0002-4908-4925]{Carsten Kramer}
\affiliation{Institut de Radioastronomie Millim\'etrique, 300 rue de la Piscine, 38406 St. Martin d’Hères, France}

\author[0000-0002-2953-7376]{Kuo Liu}
\affiliation{Max-Planck-Institut f\"{u}r Radioastronomie, Auf dem H\"{u}gel 69, D-53121, Bonn, Germany}

\author{Gabriel Paubert}
\affiliation{Institut de Radioastronomie Millim\'etrique, Avda. Divina Pastora 7, Local 20, 18012 Granada, Spain}

\author[0000-0003-0981-9664]{Miguel Sanchez-Portal}
\affiliation{Institut de Radioastronomie Millim\'etrique, Avda. Divina Pastora 7, Local 20, 18012 Granada, Spain}

\author[0000-0003-2890-9454]{Karl~F.~Schuster}
\affiliation{Institut de Radioastronomie Millim\'etrique, 300 rue de la Piscine, 38406 St. Martin d’Hères, France}


\begin{abstract}

We present the first detection of pulsations from a neutron star in the submillimeter range. The source is the magnetar XTE~J1810$-$197, observed 
with the James Clerk Maxwell Telescope (JCMT) on 2020 February 27, 2020 July 9 and 2021 May 15. XTE~J1810$-$197 is detected at 353$\,$GHz ($\uplambda\!=$0.85$\,$mm) in the three epochs, but not detected in the simultaneously-observed band at 666$\,$GHz ($\uplambda\!=$0.45$\,$mm). We measure an averaged flux density at 353$\,$GHz of 6.7$\pm$1.0, 4.0$\pm$0.6, and 1.3$\pm$0.3$\,$mJy and set 3$\sigma$ flux density upper limits at 666$\,$GHz of 11.3, 4.7 and 4.3$\,$mJy, at each of the three observing epochs, respectively. Combining close-in-time observations with the Effelsberg~100m and IRAM~30m telescopes covering non-contiguously from 6 to 225$\,$GHz (5.0$\,$cm$>\uplambda>$1.33$\,$mm), we investigate the spectral shape and frequency range of a potential spectral turn-up predicted by some pulsar radio emission models. The results demonstrate that the beamed radio emission from neutron stars can extend into the submillimeter regime, but are inconclusive on the existence and location of a potential spectral turn-up within the covered frequency range. The observed properties of the submillimeter emission resemble those of the longer wavelengths, and support a coherent mechanism for the production of pulsations at 353$\,$GHz.


\end{abstract}


\section{Introduction} \label{sec:intro}

Magnetars are a subfamily of highly-magnetized rotating neutron stars (pulsars) showing the largest inferred magnetic fields \citep[for a review on magnetar properties see e.g.,][]{kaspibelo17}.
The majority of magnetar detections are made at high energies (X$-$ and $\gamma-$ray). Nonetheless, of the 30 currently known magnetars\footnote{\url{http://www.physics.mcgill.ca/~pulsar/magnetar/main.html}} \citep{ok14}, six have been detected at radio wavelengths \citep{cam2006, cam07b, lev10, eat13, champ20, chime20_1935}. 

Magnetar radio emission shows similarities but also notable differences from the rest of the pulsar population \citep[e.g.,][]{kra07, lower21}.
One of these remarkable properties is a spectral index that can be flat or even inverted \citep{cam07b, laz08, lev10, tor15, tor17, dai19, huang2021}. As pulsars typically show steep spectra in radio \citep[$<\alpha>=-1.8$, for $S\propto\nu^{\alpha}$, e.g.,][]{mar2000}, the peculiar spectral properties of magnetars make them unique sources in which to study the emission characteristics up to very high radio frequencies.

The underlying radio emission mechanism of pulsars is still under debate \citep[e.g.,][]{melrose17, melrose21}. Some models propose a scenario in which the incoherent component of curvature emission or inverse Compton scattering produces a turn-up in the spectral energy distribution \citep{blandsch76, mich82}. The exact location of this potential turn-up
is not known, but hypothesized to occur between radio and infrared (IR) wavelengths \citep{mich82}. Identifying the frequency of such a spectral turn-up would allow for a measurement of the coherence length of the emission process, a valuable observable for testing and refining our models of pulsar magnetospheres and radiation mechanisms.

The scenario of a coherence breakdown and spectral turn-up is supported by the spectral shape of the Crab (PSR~B0531+21), Vela (PSR~B0833$-$45), and Geminga (PSR~B0633+17) pulsars.
The IR and/or optical emission of these pulsars exceeds the extrapolation from radio flux densities \citep{neugebauer69, dani11}. Furthermore, the IR and optical emission of the Crab pulsar can be explained by an incoherent process \citep{cru01}.

The first observations of pulsars in the millimeter band showed hints of spectral turn-ups around $\sim$30$-$40$\,$GHz \citep{kra96, kra97_unexpected}, with results suggesting that spectral turn-ups may not appear at the same frequency range for all pulsars \citep{morr97, loh08}. Later, observations of radio magnetars in the short millimeter range ($\sim$80$-$300$\,$GHz) showed cases where the millimeter flux density clearly exceeds the low-frequency values \citep{tor17, chu2021}. Despite the hints of potential turn-ups in the millimeter window, there is yet no satisfactory conclusion on the location of a transition frequency.

In the IR, optical, and ultraviolet bands, only about 25 neutron stars (including rotation-powered pulsars, magnetars, and X-ray dim isolated neutron stars) have been detected. In only a handful of the cases pulsations were resolved \citep[e.g.,][]{cocke69, ransom94, kermar02, kar07, dhill09}. At these wavelengths it was often difficult to identify the exact origin of the detected emission (fallback disc, and/or magnetospheric), with both thermal and non-thermal components contributing in many cases \citep[for a review see][and references therein]{mignani2011}.

The submillimeter window\footnote{We define here submillimeter window as 1$\,$mm $>\uplambda> $ 100$\,\mu$m (300$\,$GHz $\lesssim \nu \lesssim $ 3$\,$THz).} linking the short millimeter and IR bands of pulsar emission is almost totally unexplored, but has the potential to hide the spectral turn-up. Only one detection of a pulsar that extends into the submillimeter band is available to date \citep[the Vela pulsar,][]{mignani17}. Those observations, made in imaging mode with the Atacama Large Millimeter/Submillimter Array (ALMA), showed that the (sub)millimeter spectrum of Vela flattens compared to the longer-wavelength emission, and still maintains a level of coherence \citep{mignani17}.


In this paper, making use of a novel observing method with the SCUBA-2 Transition Edge Sensor (TES) camera on the James Clerk Maxwell Telescope (JCMT), we present phase-resolved observations of the magnetar XTE~J1810$-$197 in the submillimeter regime. For one epoch, we use nearly-simultaneous observations with the Effelsberg~100m and IRAM~30m telescopes to investigate the spectral shape and a potential spectral turn-up within the submillimeter band.

\section{Observations and data reduction}\label{sec:red}

\subsection{James Clerk Maxwell Telescope}

XTE~J1810$-$197 was observed on 2020 February 27, 2020 July 9, and 2021 May 15 simultaneously at central frequencies 353 and 666 GHz ($\uplambda\!=$0.85 and 0.45$\,$mm) using the SCUBA-2 camera \citep{holland13} at the JCMT, on Maunakea, Hawai'i. The effective bandwidth was 35 and 47$\,$GHz at 353 and 666$\,$GHz respectively, and the data of the bolometer arrays were sampled at $t_{\rm s}\approx 5.7\,$ms\footnote{We note that SCUBA-2 does not produce perfectly regularly sampled data. 5.7$\pm$0.6$\,$ms is the mean value of the sampling interval.}.

The observing strategy and data reduction from the JCMT were adapted to extract the potential submillimeter pulsations from the magnetar. 
With this aim, a custom small daisy map\footnote{Daisy refers to the shape of the telescope scan pattern. See Appendix \ref{app:scanpat} for details.} was prepared, intended to keep the source within one sub-array in each of the bands, minimizing interruptions to the on-source time-stream by avoiding bad bolometers and gaps between sub-arrays. Because the 0.45$\,$mm detection is most challenging, we selected a focal plane tracking location to maximize the number of good bolometers within the pattern at this wavelength. Furthermore, six of the eight sub-arrays of SCUBA-2 were disabled (those not being used), in order to achieve a small but noticeable improvement in sampling time.

Two new data analysis techniques by which pulsations can be extracted from the raw data were developed. Most straightforwardly, a new \texttt{makemap} ``cyclemap'' parameter was included within the software suite {\sc starlink}\footnote{\url{https://starlink.eao.hawaii.edu/starlink}} allowing the generation of a given number of maps corresponding to phase bins through the pulsar spin period, with the data being folded\footnote{Folding is a technique by which the data is synchronously averaged in blocks of size equal to the spin period of a pulsar.} at this period.
The second technique uses \texttt{makemap}'s diagnostic outputs to obtain a final cleaned data stream and an astrometric look-up table. From these a time series was extracted from one or more bolometers closest to the source position using a Gaussian-shaped filter. The time series was then folded with XTE~J1810$-$197's spin period, derived from the close-in-time observations from the IRAM~30m telescope (see Sec.~\ref{sec:red_30m}) and refined to the best topocentric period at the JCMT using a periodogram. An opacity extinction correction was applied by \texttt{makemap} based on measurements from the JCMT water vapor monitor, taken throughout each observation.
Both the maps and time series were calibrated to Jansky scale by multiplying the data by the recently-updated standard Flux Conversion Factors (FCF) for SCUBA-2 \citep{mairs21}.

We provide more details on the new observing strategy with SCUBA-2 and the data reduction methods for phased-resolved mapping and time domain analysis in Appendix~\ref{app:scanpat}.

\subsection{IRAM~30m}\label{sec:red_30m}

On 2020 July 10, we observed the same magnetar at central frequencies 86, 102, 138, 154, 209 and 225$\,$GHz ($\uplambda\!\simeq$3.44, 2.97, 2.17, 1.95, 1.43, and 1.33$\,$mm) with the Eight MIxer Receiver \citep[EMIR;][]{car12} at the IRAM~30m telescope, in Spain. A broadband continuum backend sampled 8$\,$GHz of bandwidth per frequency band with a sampling time of $t_{\rm s}=$100$\,\mu$s. 

The data processing consisted of a calibration to Jansky scale by Y-factor measurements on loads of known temperature, plus a correction for atmospheric opacity \citep{ckra97, pardo01}. The time series from each linear feed at each frequency band were filtered to reduce the low frequency noise by a running window of 5$\,$s that fits and subtracts a first order polynomial. A second filtering was required to remove sporadic short-duration power drop-offs of instrumental origin. The two linear polarizations were then combined to produce total intensity time series per frequency band. The time series were then folded with the spin period of XTE~J1810$-$197 obtained using the \texttt{prepfold} routine of the {\sc presto} software\footnote{\url{https://github.com/scottransom/presto}} to refine the period derived from the ephemeris given in \citet{lev19}. Finally, a correction of the flux density values at 209 and 225$\,$GHz by a factor 1.38 is applied to account for a residual pointing offset of 3.8 arcsec affecting only those frequency bands.

\subsection{Effelsberg 100m}



The Effelsberg 100m telescope in Germany observed the source on 2020 July 8 at a central frequency of 6$\,$GHz with the S45mm receiver and a dedicated pulsar backend based on two CASPER\footnote{\url{https://casper.berkeley.edu/}} ROACH2 boards, offering 4$\,$GHz of instantaneous bandwidth split into 4096 frequency channels sampled at $t_{\rm s}=131.072\,\mu$s.

The data recorded in PSRFITS search mode were first corrected for the effect of interstellar dispersion and folded modulo the spin period of the magnetar using the {\sc psrfits\_utils} tools\footnote{\url{https://github.com/gdesvignes/psrfits_utils}} and the parameters in \citet{lev19}. The averaged pulse profile was calibrated for polarization and flux density using observations of a pulsed noise diode fired on a reference source, the planetary nebula NGC7027. Finally the data was cleaned of radio frequency interference. The data post processing was done with {\sc psrchive} \citep{vanStrat12_psrchive}.

Table~\ref{tab:obstab} outlines the observing sessions.

\begin{deluxetable}{ccccccc}
\tablenum{1}
\tablecaption{Summary of observing epochs. Columns indicate: Epoch of observation, telescope, central observing frequency ($\nu_c$), instantaneous bandwidth ($\Delta \nu$), integration time on-source ($t_{\rm obs}$), and average 225-GHz zenith atmopsheric opacity during the observation ($<\!\tau_{225}\!>$). For Effelsberg, ``n/a" indicates that no opacity measurement is available. At 6$\,$GHz the atmospheric effects are negligible. 
\label{tab:obstab}}
\tablewidth{0pt}
\tablehead{
\colhead{Epoch} & \colhead{Telescope} & \colhead{$\nu_c$} & \colhead{$\Delta \nu$} & \colhead{$t_{\rm obs}$}   & $<\!\tau_{225}\!>$ \\
\colhead{(MJD)}     &                  & \colhead{(GHz)}   & \colhead{(GHz)}        & \colhead{(min)}              & (Np)         \\ 
}
\startdata
58906.639  & JCMT       & 353   & 35   & 64.4  & 0.05 \\
          &            & 666   & 47   & 64.4  & 0.05 \\
59039.296  &            & 353   & 35   & 225.4  & 0.05\\
          &            & 666   & 47   & 225.4  & 0.05\\
59349.435  &            & 353   & 35   & 226.9  & 0.04\\
          &            & 666   & 47   & 226.9  & 0.04\\
59040.958  & IRAM       & 225   & 8    & 70.2  & 0.35 \\
59040.958  &            & 209   & 8    & 70.2  & 0.35 \\
59040.872  &            & 154   & 8    & 50.2   & 0.54 \\
59040.872  &            & 138   & 8    & 50.2   & 0.54 \\
59040.872  &            & 102   & 8    & 120.4   & 0.43 \\
59040.872  &            & 86    & 8    & 120.4   & 0.43 \\
59038.995  & Effelsberg & 6  & 4    & 19.7   & n/a \\
\enddata
\tablecomments{The epoch is given here as MJD to increase the precision of the datum to $\sim$1$\,$min from the start of the observation.}
\end{deluxetable}



\section{Results}\label{sec:results}

\begin{figure*}
\includegraphics[width=\textwidth]{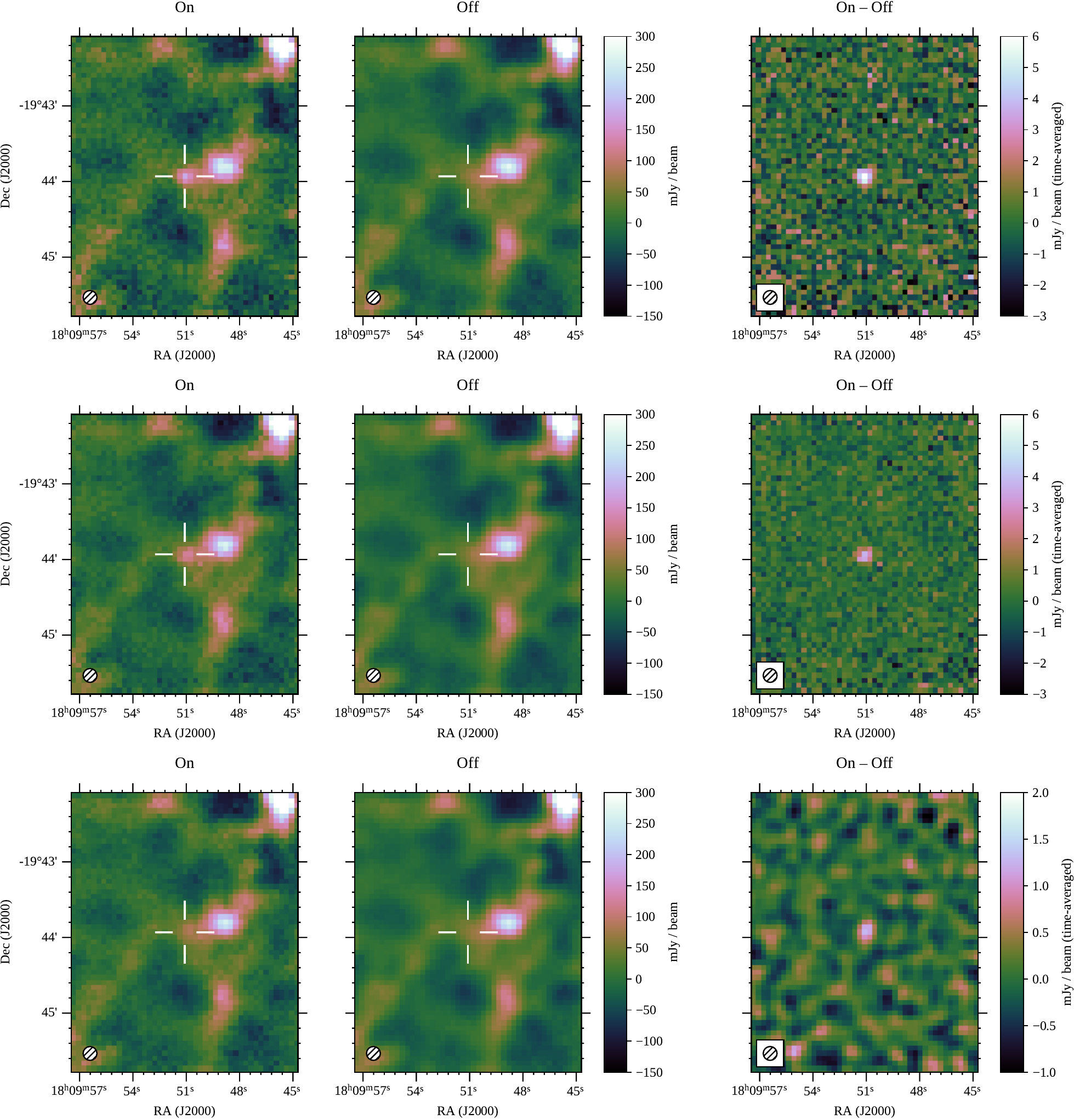}
\caption{XTE~J1810$-$197 detection with JCMT/SCUBA-2 at $\nu\!\approx$353$\,$GHz ($\uplambda=$0.85$\,$mm). The upper three panels show the results from 2020 February 27, the three middle panels from 2020 July 9, and the three bottom panels from 2021 May 15.
In each case, the total integration time was folded with a resolution of 50 maps over the spin period of the magnetar ($P\approx$5.54$\,$s), thus with a time resolution of 110.8$\,$ms. For each epoch, the panels show: (Left:) average of the three maps containing the pulsation from the neutron star, (Center:) average of 39 maps during the off-pulse time (those separated at least 4 maps from the 3 maps containing the ON signal), (Right:) resulting image after subtracting the off-pulse average map from the on-pulse average, showing the residual of the on-pulse emission from XTE~J1810$-$197. The telescope beam size is shown in the lower-left corner on each panel. For 2021 May 15, the On-Off map is smoothed  with a matched filter with primary Gaussian component of 11 arcsec to improve the visualization of the residual.}
\label{fig:maps}
\end{figure*}

\begin{figure}
\includegraphics[width=0.98\columnwidth]{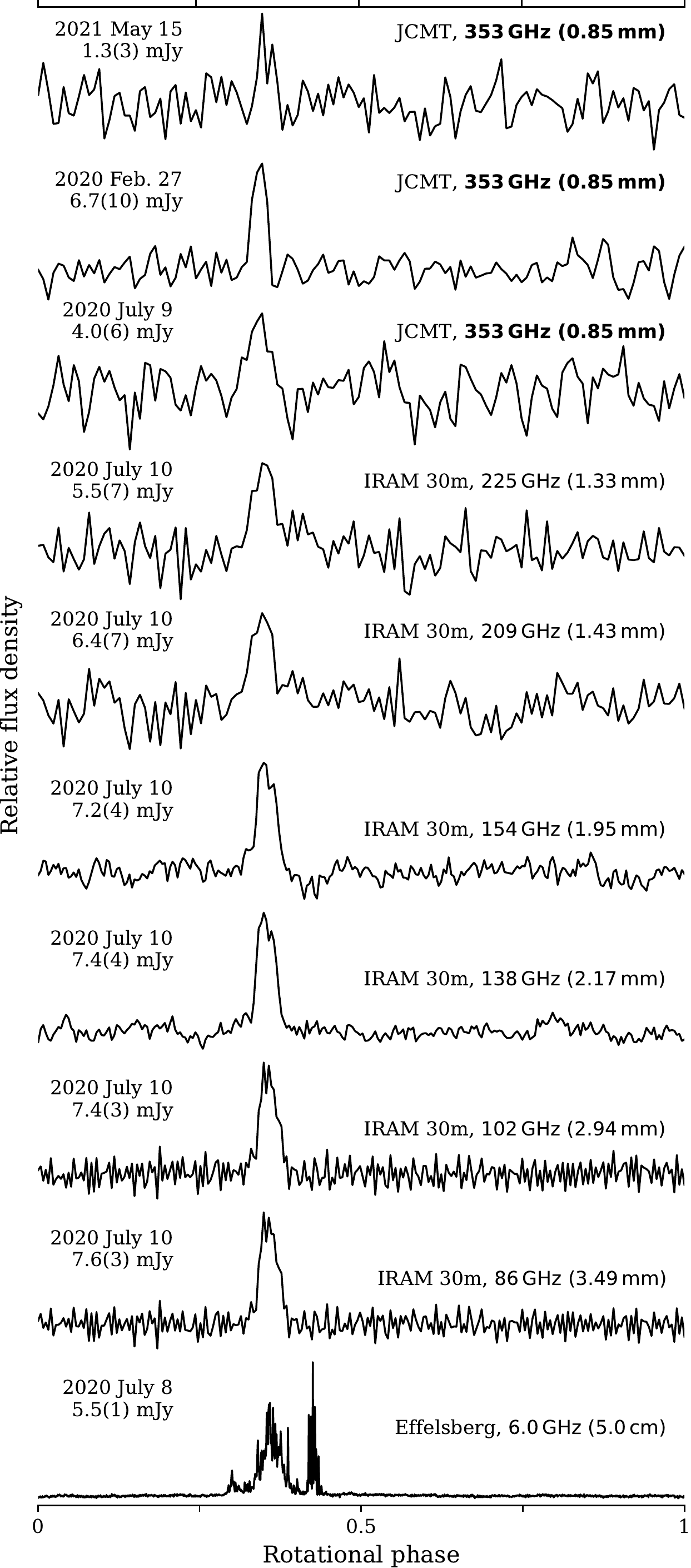}
\caption{Average pulse profiles of the magnetar XTE~J1810$-$197 at the different frequency bands. 
The top three panels show the detections at 353$\,$GHz ($\uplambda\!=$0.85$\,$mm).
Each profile includes the telescope, frequency, date of observation, and averaged flux density in the legends.
The time resolution is adapted to improve the signal-to-noise ratio, being 43.3$\,$ms (128 bins) between 209 and 353$\,$GHz, 21.6$\,$ms (256 bins) between 86 and 154$\,$GHz and 2.7$\,$ms (2048 bins) at 6$\,$GHz. The profiles were manually aligned to the peak of the main component.
\label{fig:profiles}}
\end{figure}

\begin{figure}
\includegraphics[width=1\columnwidth]{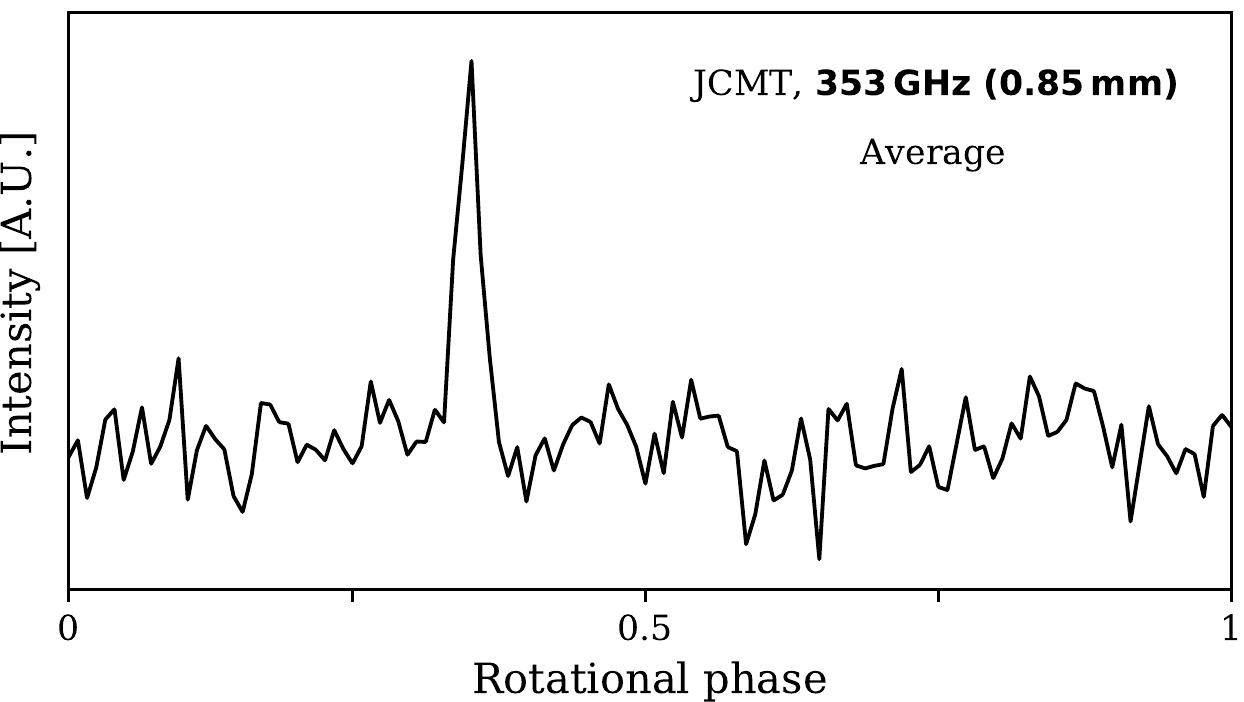}
\caption{Mean profile of the magnetar XTE~J1810$-$197 at 353$\,$GHz ($\uplambda\!=$0.85$\,$mm) after averaging the profiles from the observations of 2020 February 27, 2020 July 9 and 2021 May 15 from the JCMT.
\label{fig:submm-ave}}
\end{figure}

The magnetar XTE~J1810$-$197 was detected at 353$\,$GHz ($\uplambda\!=$0.85$\,$mm) at the three observing epochs with the JCMT, 2020 February 27, 2020 July 9, and 2021 May 15 with peak signal-to-noise ratios\footnote{Here defined as the maximum value of the average profile divided by the standard deviation of the off-pulse region.} of 8.6, 3.7, and 4.9 respectively. The measured averaged flux density (i.e., the flux density integrated over the pulse and divided by the spin period) is 6.7$\pm$1.0, 4.0$\pm$0.6, and 1.3$\pm$0.3 mJy (1$\sigma$ errors\footnote{All error figures through the paper represent 1$\sigma$ uncertainty intervals.}), for the three epochs, respectively. However, the magnetar was not detected in the simultaneously-observed band at 666$\,$GHz ($\uplambda\!=$0.45$\,$mm) in any of the epochs. From the standard deviation of an empty region around the source position in the 666-GHz maps we derive flux density upper limits of 11.3, 4.7 and 4.3$\,$mJy (for a 3$\sigma$ detection) for each of the three observing epochs, respectively. Figure~\ref{fig:maps} shows the detection maps, and the top three panels of Figure~\ref{fig:profiles} the averaged pulse profiles at 353$\,$GHz for 2020 February 27, 2020 July 9 and 2021 May 15. In Fig.~\ref{fig:submm-ave} we show the mean profile at 353$\,$GHz after averaging together the profiles from the three observing epochs at the JCMT.

On 2020 July 10, XTE~J1810$-$197 was detected at all observing frequencies between 86 and 225$\,$GHz with the IRAM~30m telescope ($\uplambda\!\simeq$3.44$-$1.33$\,$mm) with peak signal-to-noise ratios ranging from 10.3 to 19.1. The measured averaged flux densities were 7.6$\pm$0.3, 7.3$\pm$0.3, 7.4$\pm$0.4, 7.2$\pm$0.4, 6.4$\pm$0.7, and 5.5$\pm$0.7 mJy at 86, 102, 138, 154, 209 and 225$\,$GHz, respectively. At Effelsberg, the source was detected on 2020 July 8, with a more complex profile shape and a peak signal-to-noise ratio of 173, with an averaged flux density of 5.5$\pm$0.1$\,$mJy. Table~\ref{tab:FDsum} summarizes the results, and the profile shapes at the millimeter and centimeter bands are shown in Figure~\ref{fig:profiles}.

\begin{deluxetable*}{lcccccccccccc}
\tablenum{2}
\tablecaption{Averaged flux densities and spectral index for XTE~J1810$-$197. Columns show epoch of observation, averaged flux density ($S_{\nu}$) at each observing frequency, spectral index from a single power law model ($\alpha$), and spectral indices for a broken power law model ($\alpha_1$, $\alpha_2$).  A ``$-$'' symbol indicates that no observation was done at that frequency for the epoch. At 666$\,$GHz, where observations took place but with no detection of the source, 3$\sigma$ upper limits are given, indicated by the ``$<$'' symbol preceding the value. The ``$-$'' symbol in the $\alpha$, $\alpha_1$, and $\alpha_2$ columns indicates that no spectral fit is done for that epoch for the corresponding model. Values in parenthesis indicate the 1$\sigma$ error on the least significant figures.
\label{tab:FDsum}}
\tablewidth{0pt}
\tablehead{
\colhead{Epoch} & \colhead{$S_{6}$} & \colhead{$S_{86}$} & \colhead{$S_{102}$} & \colhead{$S_{138}$} & \colhead{$S_{154}$} & \colhead{$S_{209}$} & \colhead{$S_{225}$} & \colhead{$S_{353}$} & \colhead{$S_{666}$} &   $\alpha$ &   $\alpha_1$  &   $\alpha_2$ \\
\colhead{}  &   \colhead{(mJy)} & \colhead{(mJy)} & \colhead{(mJy)} & \colhead{(mJy)} & \colhead{(mJy)} & \colhead{(mJy)} & \colhead{(mJy)} & \colhead{(mJy)} &\colhead{(mJy)} & \colhead{} & \colhead{}    & \colhead{}
}
\startdata
2020 Feb. 27 & $-$      & $-$    & $-$    & $-$    & $-$    & $-$    & $-$    & 6.7(10) & $<\,$11.3 & $-$   & $-$   & $-$ \\
2020 Jul. 8  & $5.5(1)$ & $-$    & $-$    & $-$    & $-$    & $-$    & $-$    & $-$     & $-$ & $+$0.1(1)   & $-$   & $-$  \\
2020 Jul. 9  & $-$      & $-$    & $-$    & $-$    & $-$    & $-$    & $-$    & 4.0(6)  & $<\,$4.7 & $-$0.3(2)  & $-$   & $-$  \\
2020 Jul. 10 & $-$      & 7.6(3) & 7.3(3) & 7.4(4) & 7.2(4) & 6.4(7) & 5.5(7) & $-$     & $-$ & $-$0.2(2)   & $+$0.1(1)   & $-$0.7(4) \\
2021 May 15  & $-$      & $-$    & $-$    & $-$    & $-$    & $-$    & $-$    & 1.3(3)  & $<\,$4.3 & $-$  & $-$   & $-$ \\
\enddata
\tablecomments{2020 July 10 is the only epoch with observations at several frequencies within a single observing session. The spectral indices given for 2020 July 9 and 2020 July 8 and those from the broken power law arise from combinations of flux density values from different days. See Fig.~\ref{fig:spectrum} and Sec.~\ref{sec:results} for details.}
\end{deluxetable*}

\begin{figure}
\includegraphics[width=1\columnwidth]{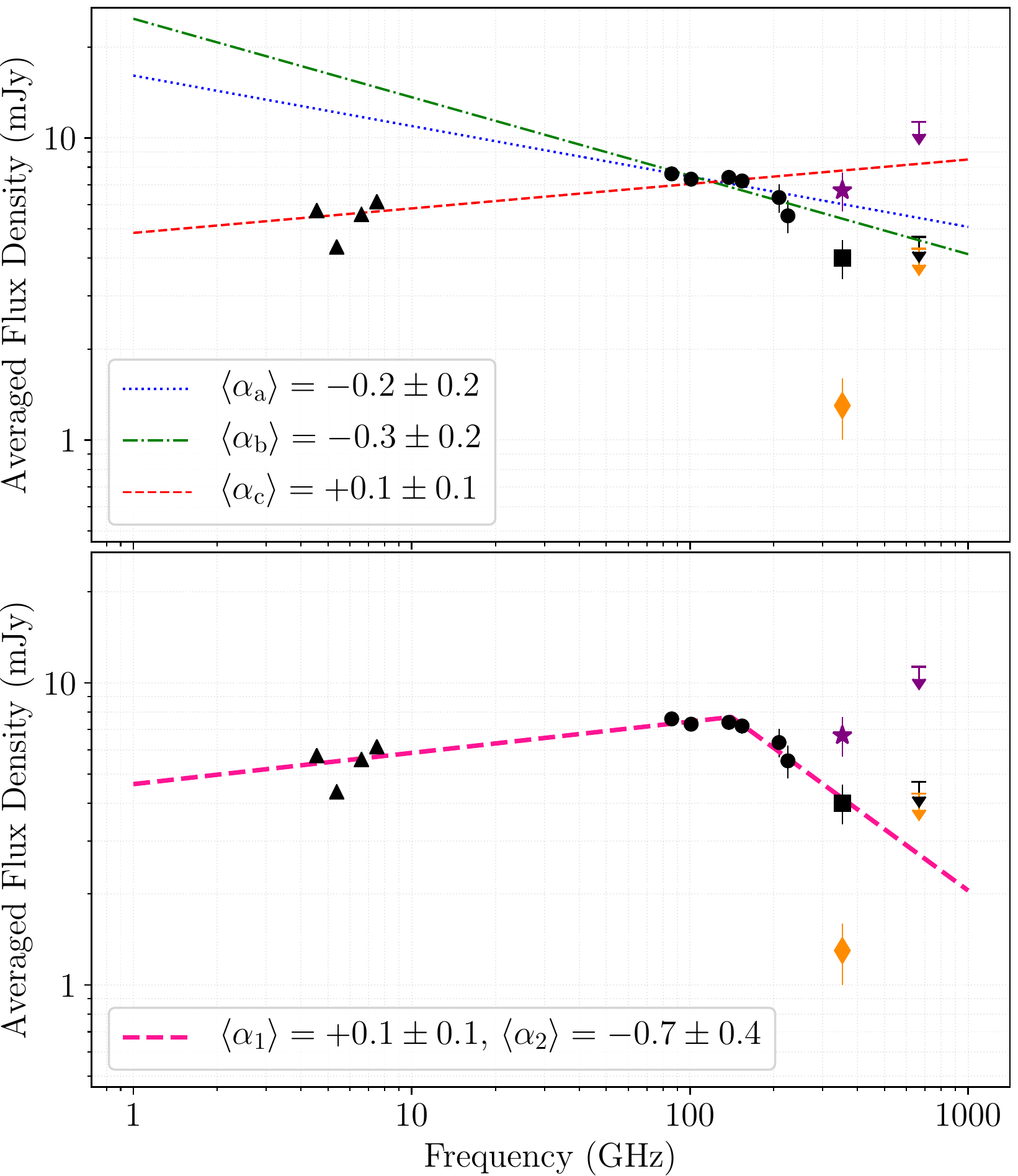}
\caption{Spectral fits for XTE~J1810$-$197. The upper panel shows single power law fits, and the bottom panel a broken power law. In both panels, the triangle markers show the centimeter averaged flux densities from Effelsberg at 4.5, 5.5, 6.5 and 7.5$\,$GHz on 2020 July 8, the circles show the millimeter averaged flux densities from the IRAM~30m telescope at 86, 102, 138, 154, 209 and 225$\,$GHz on 2020 July 10, and the black square the averaged flux density from the detection from JCMT at 353$\,$GHz on 2020 July 9. The purple star and orange diamond represent the averaged flux density measured at 353$\,$GHz on 2020 February 27 and 2021 May 15 for comparison, but they are not used for the power law fits. 1$\sigma$ error bars are shown, but are often smaller than the size of the symbols. The black, purple and orange down-pointing arrows at 666$\,$GHz mark the 3$\sigma$ upper flux density limits for our observation from the JCMT on 2020 July 9, 2020 February 27 and 2021 May 15, respectively. The top panel shows three different single power law fits: The blue dotted line using only the IRAM~30m data, the dotted-dashed green line is the fit to a combination millimeter and submillimeter data, and the dashed red line shows the single power fit to all data from 2020 July combined.
See Sec.~\ref{sec:results},~\ref{sec:dis} and Table~\ref{tab:FDsum} for details. 
\label{fig:spectrum}}
\end{figure}

With the caveat of the known short-term variability from radio magnetars, including XTE~J1810$-$197 \citep[e.g.,][]{cam07b, tor15, tor20} and the potential impact of interstellar scintillation \citep{cam2006, laz08}, we try to investigate the spectral shape of XTE~J1810$-$197. Firstly, we use only data from the IRAM~30m telescope on 2020 July 10, that cover several frequencies ($\nu\in[86,225]\,$GHz) during a single observing session. We consider this dataset as less affected by intrinsic intensity variations. A single power law fit through error-weighed least squares yields $\alpha_{\rm a}=-$0.2$\pm$0.2, for $S_{\nu}\propto\nu^{\alpha}$. Including the data point at 353$\,$GHz from the JCMT on 2020 July 9, the fit yields $\alpha_{\rm b}=-$0.3$\pm$0.2, which is consistent within the uncertainties. We note however that the extrapolation to centimeter wavelengths in any of these cases gives a significantly higher flux density at $\sim$6$\,$GHz than is measured. We then fit the combined dataset with all the flux density measurements of 2020 July 8, 9 and 10 from the three telescopes. Taking advantage of the large fractional bandwidth ($\Delta \nu / \nu$) of the Effelsberg observations, its total bandwidth was split into four 1-GHz chunks centered at 4.5, 5.5, 6.5 and 7.5$\,$GHz. The result in this case yields $\alpha_{\rm c}=+$0.1$\pm$0.1. Nonetheless, in this latter case the data above $\sim$200$\,$GHz are poorly fit. Finally, we fit a broken power law model \citep[see e.g.,][]{jank18}, again to combined data of 2020 July 8, 9 and 10 from the three telescopes. This model can account for the apparent spectral break and fits the spectrum better than a single power law. The yielded spectral indices are $\alpha_1=+$0.1$\pm$0.1 and $\alpha_2=-$0.7$\pm$0.4, respectively for the spectral ranges before and after the frequency of the break, which is best-fit at a frequency of 141$\pm$1$\,$GHz. Figure~\ref{fig:spectrum} presents the measured spectrum of XTE~J1810$-$197 and compares the spectral fits.

\section{Discussion and conclusion}\label{sec:dis}

\subsection{Variability, pulse morphology, and spectral shape}\label{sec:varpulsespec}

The observations of XTE~J1810$-$197 extending for the first time into the submillimeter regime of magnetar radio emission show similar characteristics in the submillimeter emission to those at lower frequencies. There is a clear variability, with a decrease at 353$\,$GHz of about 40\% in the flux density between 2020 February 27 and 2020 July 9, and a factor $\sim$5 between 2020 February 27 and 2021 May 15. Moreover, the average pulse morphology at 353$\,$GHz resembles to a high degree that of the millimeter emission between 86 and 225$\,$GHz, with a single dominant component. The average pulse at 6$\,$GHz is more complex, but shows a component equivalent to the one observed at the higher frequencies.

The radio spectrum over the full frequency range covered is best described by a broken power law with a break at a frequency of $\approx$141$\,$GHz. The spectral indices both before and after the break are broadly consistent with previous measurements reporting a variable spectral index from XTE~J1810$-$197, during the 2003 and 2018 outbursts, of about $-$1.1$< \alpha <$+0.3 \citep{cam07b, laz08, dai19, tor20}. The submillimeter pulsed emission observed on 2020 July 9 is not brighter than the emission at lower frequencies from close-in-time observations, and therefore does not show evidence for a turn-up in the emission in the range up to 353$\,$GHz. We remark however that the fact that the measurements at the centimeter, millimeter, and submillimeter wavelengths are not fully simultaneous adds some uncertainty to this conclusion. We do not use the 2020 February 27 or 2021 May 15 detections at 353$\,$GHz to derive conclusions on a potential spectral turn-up because of the strong variability of the source, which invalidates the comparison of flux densities measured at epochs separated by such a long time.

Various reasons could explain why no single spectral index fits well all the data points. Radio magnetars (including XTE~J1810$-$197) show significant intensity variations on short time scales, even intraday \citep[e.g.][]{cam07b, tor15, tor20}, and this can affect the fit when combining data that, even when very close in time, are not fully simultaneous. Furthermore, the potential impact of interstellar scintillation cannot be neglected, in particular when a large frequency range is covered, as in this case. Weak and strong scintillation effects can have different influence at different frequencies \citep[see e.g.,][]{lorkra04}.

Notably, the profile at 6$\,$GHz shows more components than that at the higher frequencies. A similar profile evolution with frequency was observed for the magnetar SGR~J1745$-$2900 \citep{tor15}, and could indicate the we are not sampling through the same emission region at all wavelengths, making the comparison of total flux density between frequencies imprecise. The lack of several components seen at the higher frequencies can also mean that the spectral index of the different components is different\footnote{In fact, the difference in flux density observed for the four 1-GHz subbands in the Effelsberg data (see Fig.~\ref{fig:spectrum}) are due to an uneven spectral evolution of the different components.}. However, if we would select only the broad component from 6$\,$GHz to fit a single power law, the measured spectral index $\alpha_{\rm c}$ would be even more inverted and would not improve the fit to the points above 200$\,$GHz. Full-polarization information at the millimeter and submillimeter wavelengths would enable a better understanding of the viewing geometry \citep[e.g.,][]{liu21}, and a comparison with the measurements at centimeter wavelengths \citep{kra07, camrey07_1810pol, dai19}. The polarization degree and position angle could help to identify equivalent components among frequency bands.


For the highest frequencies, given the small size of the telescope beams (e.g., $\theta\approx$11.8, 11.0, 14.0, 7.4$\arcsec$ for our observations at 209, 225, 353 and 666$\,$GHz, respectively\footnote{$\theta$ denotes the Half Power Beamwidth (HPBW), i.e., the angular separation in which the received power decreases by 50\%.}), even very small pointing and focus errors may translate into non-negligible intensity underestimations. The data points at 209, 225 and 353$\,$GHz show a lower flux density compared to the other values between 86 and 154$\,$GHz (see Table~\ref{tab:FDsum} and Fig.~\ref{fig:spectrum}). Our data processing took into account potential pointing offsets (see Sec.~\ref{sec:red_30m} and Appendix~\ref{app:scanpat}), so we do not consider this the main source of the observed spectral deviation from the power law.

Finally, a plausible explanation is simply that the radio spectrum of XTE~J1810$-$197 is not well described by a single power law over wide frequency ranges, but shows a more complex shape. A relatively high number of pulsars show spectral shapes deviating from single power laws \cite[see][]{mar2000, jank18}. Interestingly, spectra showing (or supporting) spectral turn-overs have been reported in previous observations of XTE~J1810$-$197 \citep{dai19}, and in another four radio magnetars \citep{camrey08, tor17, chu2021, lower21}. 
As shown in Fig.~\ref{fig:spectrum}, this is the case again for our observations of XTE~J1810$-$197, where a broken power law fits the spectrum better than single power laws. These results indicate that radio magnetar spectra may often show features like spectral turn-overs, and that these turn-overs may appear at different frequencies and even change shape over time along with the frequency-dependent intensity variations.

As more observations of pulsars and magnetars become available covering large frequency ranges in the radio band, it is clear that the region between $\sim$10$-$70$\,$GHz is under-sampled, but can be key to fully understand the spectral energy distribution of neutron star radio emission. The same lack of information between $\sim$8 and 86$\,$GHz for the observations of XTE~J1810$-$197 presented here may well be hiding remarkable spectral features in this region.

\vspace*{0.5cm}

\subsection{Brightness temperature and coherence}

We can infer more information on the emission origin from the brightness temperature ($T_{\rm B}$) of the pulsations. $T_{\rm B}$ has an inverse quadratic dependence on the pulse width \citep[e.g.,][]{lorkra04}, and so it should ideally be derived from resolved single pulses, or at least single pulses detected with high time resolution. Up to the millimeter wavelengths, the radio emission mechanism of XTE~J1810$-$197 requires coherence to explain the high $T_{\rm B}$ values \citep{tor20}. For the submillimeter emission that we observed here for the first time, we unfortunately did not detect individual pulsations. 

One way to derive a lower limit for $T_{\rm B}$ at 353$\,$GHz is to assume that the shortest single pulsations are equal to the duration of the average pulse. The duty cycle of the average pulse of XTE~J1810$-$197 at 353$\,$GHz is about 6\%, that corresponds to a $\Delta t\approx$332$\,$ms for the spin period of $P$=5.54$\,$s. For a distance to XTE~J1810$-$197 of $D=$2.5$\,$kpc \citep{ding20}, the minimum brightness temperature to produce the flux density of 6.7$\,$mJy observed on 2020 February 27 is $T_{\rm B}\simeq 3\cdot10^{12}\,$K. A $T_{\rm B} \sim 10^{12}\,$K is in the limit where incoherent emission alone may explain the high brightness temperature \citep[see e.g.,][]{singal09}. Nevertheless, both XTE~J1810$-$197 and other radio magnetars usually show radio emission consisting mainly of very short pulses of the order of $\sim$1$\,$ms within (and outside) the main pulse region \citep[e.g.,][]{cam2006, lev12, spi14, whar2019, maan19}. This narrow sub-pulse characteristic extends for XTE~J1810$-$197 even to the millimeter-wavelength radiation \citep{cam07b}. 

If we accept a less restrictive assumption for the minimum duration of single pulsations of $\Delta t\!\approx$40$\,$ms, by similarity with the millimeter-wavelength single-pulse emission recently detected \citep{tor20}, the brightness temperature would reach $T_{\rm B}\sim2\cdot 10^{14}\,$K, requiring some coherence amplifying the emission. As it is likely that the single pulse emission at submillimeter wavelengths is narrower than the average profile, we consider the second scenario more plausible, in which the submillimeter emission still requires a coherent mechanism. Nonetheless, the detection of single pulses with high enough time resolution in the submillimeter band will be needed to obtain a conclusive result.

\subsection{Emission at 666 GHz and turn-up location}

If we extrapolate the power law with $\alpha_{2}$ (see Sec.~\ref{sec:results}), the flux density at 666$\,$GHz from XTE~J1810$-$197 would be about $2.7_{-1.2}^{+2.3}\,$mJy (1$\sigma$ confidence level). This value is below our detection limit of 4.7$\,$mJy for a 3$\sigma$ detection, but consistent within errors. This upper limit does not allow us to set further constrains on the potential flux density at 666$\,$GHz.
For the 2020 February 27 data, the 3$\sigma$ upper flux density limit of our observations is significantly higher than in 2020 July 9 and 2021 May 9 due to a shorter integration time and a larger impact of atmospheric opacity in this epoch. In 2021 May 15, the magnetar displayed weaker emission at submillimeter wavelengths. Thus both cases are even less constraining and we do not use them to extract any conclusion or constraint on the submillimeter spectral shape. In any case, emission from XTE~J1810$-$197 at 666~GHz cannot be ruled out by our results, which encourage higher-sensitivity observations for a more stringent limit at 666$\,$GHz. 

Our non-detections at 666$\,$GHz on 2020 July 9 are however indicative of a lack of spectral turn-up between 353 and 666$\,$GHz, that would have made the 666-GHz emission brighter than at 353$\,$GHz, and thus potentially detectable even with the available sensitivity. This result suggests that if the spectral turn-up exists, and it is above the millimeter band, it probably occurs at even higher frequencies, approaching the IR band or within it. XTE~J1810$-$197 has a known IR counterpart \citep{isra04} showing variability like that observed in the radio band \citep{testa08}. The available IR data are from the previous outburst phase of magnetar in the 2000s, and therefore not directly comparable with our current observations. The multi-wavelength variability of XTE~J1810$-$197 requires simultaneous radio and IR observations to derive strong conclusions on the possible relationship between radio and IR radiation and the spectrum shape.

A similar conclusion on the potential location of a spectral turn-up was obtained by \citet{mignani17} for the Vela pulsar. We remark nonetheless that our results and those from \citet{mignani17} cover two different types of pulsars (a magnetar and a young rotation-powered pulsar), so we cannot discard the idea that the magnetospheric characteristics may differ between the two objects, at least during a magnetar outburst and active phase \citep[e.g.,][]{belo09}. Therefore, the turn-up spectral region might be different for the two pulsars, as is, for example, the spectral index of the coherent radio emission for the two neutron stars. In addition, and given the strong known variability of radio magnetars, we should not discard the possibility of a variable turn-up spectral position in XTE~J1810$-$197. Repeated observations of the source are therefore encouraged to strengthen the constraints on the potential turn-up region.

Finally, we do not aim here to provide a physical explanation for the spectral turnover at 141$\,$GHz, to a large extent due to the different sources of potential variability in the measured flux densities of XTE~J1810$-$197 outlined in Sec.~\ref{sec:varpulsespec}. Nevertheless, several pulsar models include a potential physical interpretation for a spectral break and a decrease in the emission intensity, which are usually related to a loss of coherency under certain circumstances \citep[see e.g.,][and references therein]{sie73, belo07}.
If the spectral break at 141 GHz and observed decrease in the intensity of XTE~J1810$-$197's coherent radio emission is indeed due to a loss of coherency of the underlying emission process, this would support a transition frequency to incoherent emission that lies above our observed frequencies and the idea that we may be approaching the spectral range in which a turn-up may be observable.

\subsection{Conclusions}

The magnetar XTE~J1810$-$197 can emit submillimeter-wavelength beamed emission, detected as pulsations in our observations with the JCMT. Given the similarities of the submillimetre emission properties (variability, pulse morphology, high $T_{\rm B}$) with those at longer wavelengths, and with the submillimeter pulsations supporting a link with the co-rotating beamed magnetospheric emission, we conclude that the pulsed submillimeter emission detected from XTE~J1810$-$197 can be explained as arising from the same mechanism producing the longer-wavelength radio emission. Our observations did not show a spectral turn-up in the emission of XTE~J1810$-$197 up to 666~GHz, and suggest that the location of such a potential turn-up is probably at even higher frequencies.





\vspace*{1.5cm}

\section{Summary and Outlook}\label{sec:conc}

The high sensitivity of the SCUBA-2 camera at the JCMT together with its fast sampling capability and a novel observing and data analysis technique enabled the first detection of pulsations from a neutron star in the submillimeter band. We detected the radio magnetar XTE~J1810$-$197 in three observing epochs at $\nu\approx$353$\,$GHz ($\uplambda=$0.85$\,$mm), while no detections were achieved in the simultaneously-observed band at $\nu\approx$666$\,$GHz ($\uplambda=$0.45$\,$mm). These detections set a new record in the detection of pulsations from neutron stars in the radio band \citep[cf.][]{tor17} and demonstrate that the beamed radio emission from these objects can extend into the submillimeter range.

The properties of the emission at 353$\,$GHz resemble those of the lower radio frequencies. XTE~J1810$-$197 shows a pulse profile similar to the one observed in the short millimeter band and the broad component at centimeter wavelengths. 
Furthermore, the submillimeter emission is variable, and a lower limit for the brightness temperature of the pulses suggests still a level of coherence. These characteristics strengthen a magnetospheric origin of the submillimeter emission, as opposed to for example emission from a fallback disc, while at the same time relate the submillimeter pulses to the same coherent mechanism producing the lower-frequency emission.

The spectrum over the full observed frequency range is relatively flat and best described by a broken power law with a break frequency at about 141$\,$GHz. No clear hint of turn-up is found in the spectrum up to 666$\,$GHz. This supports the idea that the region of a potential spectral turn-up is closer to the infrared band, or above. Other remarkable spectral turn-overs in the radio spectrum of XTE~J1810$-$197 could however be hiding in the region $\sim$10$-$80$\,$GHz, not covered by the observations presented here.

The ability of the SCUBA-2 camera to simultaneously observe two frequency bands is a great advantage for testing emission models in magnetars, because of the strong short-term intensity and spectral shape variability of these objects. A similar advantage is offered by EMIR at the IRAM 30m telescope, capable of observing four separate frequencies simultaneously. In order to correctly measure the spectral energy distribution of magnetars, the biggest challenge is probably the need to obtain fully-simultaneous observations covering large frequency ranges. The strong variability of XTE~J1810$-$197, the lack of a clear identification of a spectral turn-up in its radio spectrum, and the confirmation now of the existence of pulsations in the submillimeter band, encourage follow-up observations to track the variability and evolution of its submillimeter emission. 

The advent of pulsar observations with ALMA \citep[see][]{liu19} and new high-sensitivity receivers covering the $\sim$10$-$70$\,$GHz range will certainly help to increase the number of detected neutron stars at high radio frequencies. This will help us to complete the picture of the radio spectrum from neutron stars and, in combination with the increasing capabilities of single-dish (sub)millimeter facilities and IR/Optical telescopes to detect and study pulsars, finally uncover the location of the spectral turn-up and the connection between coherent and incoherent emission in pulsar radiation.


\acknowledgments
We thank the anonymous referee for useful comments helping us to improve the manuscript. Pablo Torne is deeply grateful to the East Asian Observatory staff for their fantastic support during his visiting stay in Hilo, and thanks Prof. Dr. Paul Ho for granting Director's Discretionary Time for the
initial
JCMT observations. The JCMT data are from projects M19DB006 and M21AP032. 
The James Clerk Maxwell Telescope is operated by the East Asian Observatory on behalf of The National Astronomical Observatory of Japan; Academia Sinica Institute of Astronomy and Astrophysics; the Korea Astronomy and Space Science Institute; Center for Astronomical Mega-Science (as well as the National Key R\&D Program of China with No. 2017YFA0402700). Additional funding support is provided by the Science and Technology Facilities Council of the United Kingdom and participating universities in the United Kingdom and Canada. Additional funds for the construction of SCUBA-2 were provided by the Canada Foundation for Innovation. This work is partly based on observations carried out with the IRAM 30m telescope (under project 183-19). IRAM is supported by INSU/CNRS (France), MPG (Germany) and IGN (Spain). This work is also partly based on observations with the 100-m telescope of the MPIfR (Max-Planck-Institut für Radioastronomie) at Effelsberg. Financial support by the European Research Council for the ERC SynergyGrant BlackHoleCam (ERC-2013-SyG, GrantAgreement no. 610058) is gratefully acknowledged. Ralph P. Eatough is a `FAST Distinguished Young Researcher' under the `Cultivation Project for FAST Scientific Payoff and Research Achievement of the Center for Astronomical Mega-Science, Chinese
Academy of Sciences (CAMS-CAS)'. This research has made use of NASA's Astrophysics Data System.

%

\vspace{5mm}
\facilities{JCMT, IRAM:30m, Effelsberg}



\software{ {\sc starlink} \citep{currie14}, {\sc presto} \citep{ransom2001}, {\sc dspsr} \citep{vanStrat10}, {\sc psrchive} \citep{vanStrat12_psrchive}, {\sc psrfits\_utils} (https://github.com/gdesvignes/psrfits\_utils), {\sc numpy} \citep{numpy_paper}, {\sc scipy} \citep{scipy_paper}, {\sc matplotlib} \citep{matplotlib_paper}, {\sc astropy} \citep{astropy:2013}, {\sc aplpy} \citep{aplpy_paper} }



\appendix
\restartappendixnumbering

\section{Additional information on JCMT/SCUBA-2 scan pattern and data reduction}\label{app:scanpat}

\subsection{Custom small daisy map}

\begin{figure*}
\includegraphics[width=\textwidth]{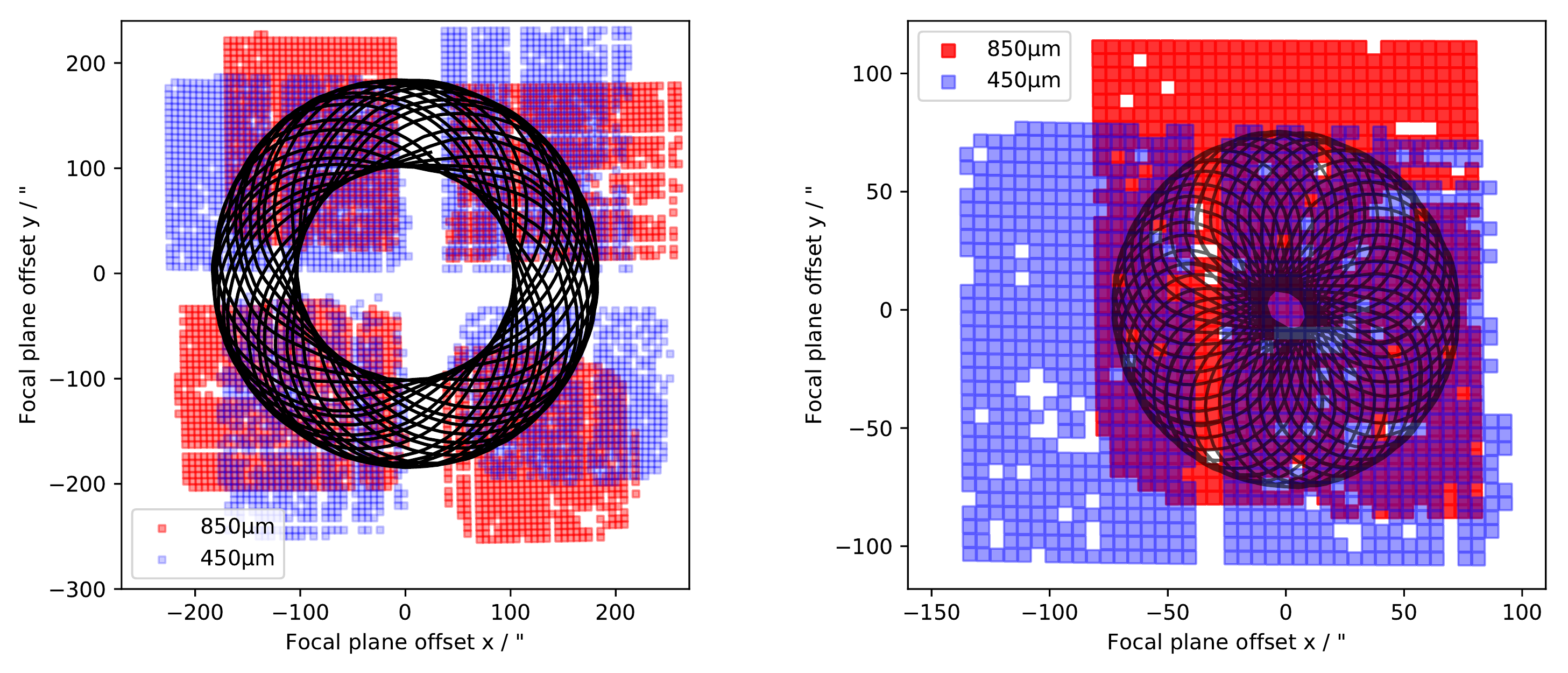}
\caption{Comparison of the standard (Left panel) and custom (Right panel) SCUBA-2 daisy maps. The red and blue squares represent the position on the focal plane of the good bolometers in the 850 and 450$\,\mu$m bands, respectively. The black line shows the position of the center of the target during a $\sim$2$\,$min scan in each of the two scan patterns. The standard daisy pattern utilizes all the eight arrays (four at each wavelength band 850 and 450$\,\mu$m) to cover a large field of view. The custom smaller daisy pattern was adapted to minimize interruptions of the time-streams, i.e., to avoid as much as possible bad bolometers and obtain the maximum number of data points for a point-like source like the magnetar. The calibration of the data was verified to be consistent between the two observing modes within 1.7\%.
\label{fig:scanpatterns}}
\end{figure*}

Conventional SCUBA-2 observing modes involve scanning the array fairly rapidly in order to allow the astronomical signal to be separated from undesired signals such as instrumental $1/f$ noise. For observations of small fields, a ``constant velocity daisy'' scan pattern is used. The telescope moves at a constant speed while making a number of loops around the target position, with the standard pattern parameters having been optimized to give uniform coverage over a three arcminute diameter \citep{holland13}. 
The 850 and 450$\,\mu$m arrays view the sky simultaneously via a dichroic beam-splitter, with each consisting of four sub-arrays separated by gaps. As shown in the left panel of Fig.~\ref{fig:scanpatterns}, the standard scan pattern is centered in the focal plane such that the target position moves between all of the sub-arrays.

In order to minimize interruptions to the time-stream, a custom daisy scan pattern was designed for the observations from the JCMT presented in this paper, to keep the target within a single pair of overlapping sub-arrays as shown in the right panel of Fig.~\ref{fig:scanpatterns}. The location within the focal plane was chosen to maximize the number of good bolometers at 450$\,\mu$m, since detection was expected to be most challenging at this wavelength. The main goal of this custom setup was to minimize the number of bad bolometers within the scan pattern, maximizing therefore the number of data points acquired on the point-like source to avoid or reduce a potential miss of pulsations from the magnetar by being on a bad bolometer when a pulse arrived to the telescope. The other sub-arrays were disabled during our observations to minimize delays during the data acquisition process --- this reduced the average sample time slightly, e.g. from $\sim 5.93\pm1.4$ to $\sim 5.66\pm0.6\,$ms.

To check that the custom scan pattern with the unused sub-arrays disabled did not affect the calibration of measured fluxes, we performed consecutive flux calibration observations on the source CRL2688 with the standard and custom patterns and found a difference of only 1.7\% --- well within the typical uncertainty for SCUBA-2 observations of 7\% \citep{mairs21}.

\subsection{Data reduction}

The data were reduced using \texttt{makemap}, part of the \textsc{starlink} package \textsc{smurf}. From initial reductions of each scan, we estimated relative pointing offsets between them of typically $\sim 2.5$ arcsec. These offsets were applied to subsequent reductions in order to align the resulting images.

The \texttt{makemap} configuration parameters were adjusted to disable down-sampling of the raw data and use Principal Component Analysis (PCA) for improved subtraction of correlated background signals. Additional diagnostic parameters enabled writing of the final cleaned time series and astrometric look-up table to files. These products were used to extract a time series for the source as it was scanned around the sub-array, applying a Gaussian weighting to each bolometer value based on its distance from the source location.

A new option ``cyclemap'' was added to \texttt{makemap} to produce a sequence of maps corresponding to phase bins through the magnetar's spin period, with the cleaned raw data being folded into these bins at the end of the data reduction process. These maps were co-added to produce the on-pulse and off-pulse maps which were subtracted to isolate the pulsar emission. Finally, the standard SCUBA-2 calibration was applied using the PICARD post-processor, part of the ORAC-DR pipeline \citep{oracdr}.


\bibliography{1810_SCUBA2}{}
\bibliographystyle{aasjournal}



\end{document}